\documentclass[showpacs,amsmath,amssymb,aps,twocolumn,longbibliography,pra]{revtex4-2}
\usepackage{graphicx}
\usepackage[colorlinks=true,citecolor=blue,linkcolor=magenta]{hyperref}
\usepackage[usenames]{color}
\usepackage{relsize}
\usepackage[english]{babel}
\usepackage{enumerate}
\usepackage{bbm}
\usepackage{xcolor}
\usepackage{booktabs}

\newcommand{\nocontentsline}[3]{}
\newcommand{\tocless}[2]{\bgroup\let\addcontentsline=\nocontentsline#1{#2}\egroup}

\newcommand {\grsim} {\ {\raise-.5ex\hbox{$\buildrel>\over\sim$}}\ }
\newcommand {\lessim} {\ {\raise-.5ex\hbox{$\buildrel<\over\sim$}}\ }

\newcommand{\nm}{\text{ nm}}
\newcommand{\ms}{\text{ ms}}
\newcommand{\mm}{\text{ mm}}
\newcommand{\um}{\mu\text{m }}

\usepackage{xcolor}
\usepackage{siunitx}

\newcommand{\RN}[1]{%
  \textup{\uppercase\expandafter{\romannumeral#1}}%
}

\usepackage{xr}
\begin{document}

\title{Optical tweezer generation using automated alignment and adaptive optics}
\author{Bharath Hebbe Madhusudhana}
\author{Katarzyna Krzyzanowska}
\author{Malcolm Boshier}
\affiliation{MPA-Quantum, Los Alamos National Laboratory, Los Alamos, NM-87544, United States}
\date{\today}

\begin{abstract}
Recent progress in quantum technologies with ultracold atoms has been propelled by spatially fine-tuned control of lasers and diffraction-limited imaging. The state-of-the-art precision of optical alignment to achieve this fine-tuning is reaching the limits of manual control.   Here, we show how to automate this process. One of the elementary techniques of manual alignment of optics is \textit{cross-walking} of laser beams. Here, we generalize this technique to \textit{multi-variable cross-walking}. Mathematically, this is a variant of the well-known Alternating Minimization (AM) algorithm in convex optimization and is closely related to the Gauss-Seidel algorithm. Therefore, we refer to our multi-variable cross-walking algorithm as the \textit{modified AM algorithm}. While cross-walking more than two variables manually is challenging, one can do this easily for machine-controlled variables. We apply this algorithm to mechanically align high numerical aperture (NA) objectives and show that we can produce high-quality diffraction-limited tweezers and point spread functions (PSF).  After a rudimentary coarse alignment, the algorithm takes about $1$ hour to align the optics to produce high-quality tweezers. Moreover, we use the same algorithm to optimize the shape of a deformable mirror along with the mechanical variables and show that it can be used to correct for optical aberrations produced, for example, by glass thickness when producing tweezers and imaging point sources. The shape of the deformable mirror is parametrized using the first $14$ non-trivial Zernike polynomials, and the corresponding coefficients are optimized together with the mechanical alignment variables. We show PSF with a Strehl ratio close to $1$ and tweezers with a Strehl ratio $>0.8$. The algorithm demonstrates exceptional robustness, effectively operating in the presence of significant mechanical fluctuations induced by a noisy environment.
    
\end{abstract}

\maketitle

\tocless\section{Introduction} 
Highly controllable quantum systems continue to make significant advances in quantum technologies, drawing attention to the challenge of scaling them up to a larger number of qubits. Two prominent platforms for this endeavor are trapped neutral atoms~\cite{Ebadi_2021, Bernien_2017, kaufman2021quantum} and trapped ions~\cite{Blatt2012, RevModPhys.93.025001, Zhang_2017}. Scaling up in both of these platforms involves creating precise, controllable, and intricate optical patterns, such as an array of optical tweezers. This necessity has spurred the exploration of opto-mechanical and opto-electronic devices, including multi-channel Acousto-Optic Modulators, Acousto-Optic Deflectors, Spatial Light Modulators, and Digital Micromirror Devices. The patterns produced by these devices are then projected using high numerical aperture (NA) objectives.

One of the challenges in developing this technology is the growing demand for precise mechanical alignment of optical systems (e.g., objectives) and the increasing number of free parameters in these devices that need optimization, often coupled with mechanical variables. Automating the optical alignment and optimizing device parameters is a potential solution to address this challenge. Furthermore, automation is crucial for scaling up and modularizing experimental platforms in general.

A common technique for automation in optics is the Nelder-Mead algorithm and its variants~\cite{ASSIMI2018127, 4026448}, often employed for automating the coupling of fiber optics~\cite{1288283}. However, these algorithms may be unsuitable in the presence of large fluctuations in both control and feedback parameters. Moreover, the control parameters are frequently not in a closed loop.

\begin{figure}[ht!]
    \includegraphics[scale=0.37]{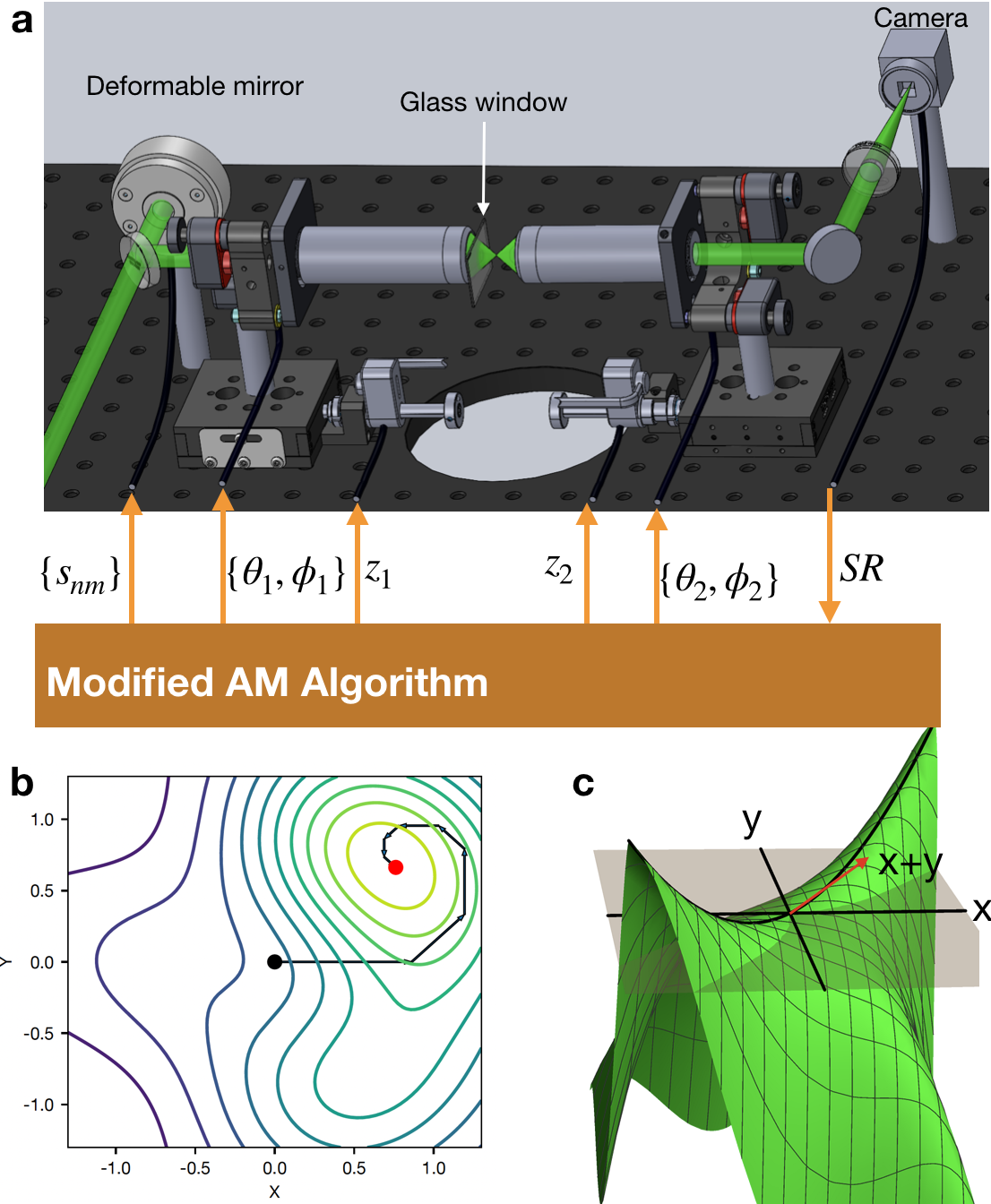}
    \caption{\textbf{Automated alignment:} \textbf{a} Basic tweezer generation apparatus, with motorized controls of the orientation of the objectives, a glass window between them and a deformable mirror. We develop an algorithm to automate the optimization of the orientation of the objectives and variables of the deformable mirror, based on multi-variable cross walking. \textbf{b} An example of the working of the modified AM algorithm for two variables. \textbf{c} The modified AM algorithm works better than the AM algorithm at saddle points.  }
    \label{Fig1}
\end{figure}

Here, we have developed an algorithm for the automation of optical alignment that exhibits robustness to experimental noise and systematics in the control and feedback parameters. This algorithm draws inspiration from the elementary cross-walking technique employed in the manual alignment of lasers (see Fig.~\ref{Fig1}b).

We generalise it to a multi-variable cross walking and show that it is related to the Gauss-Siedel algorithm in linear algebra. Moreover, in a special case, our algorithm reduces to the standard alternating minimization (AM) algorithm. We refer to our algorithm as the modified AM algorithm. We show experimentally, that this algorithm can be used to produce very high quality diffraction limited tweezers and point spread functions. Moreover, we show that the algorithm can be used to optimize a deformable mirror, along with the mechanical alignment variables, when used to produce a tweezer. We show that the algorithm is extremely robust, and works on systems built on non-vibration isolated tables, despite the presence of pronounced mechanical fluctuations arising from a noisy environment. 

One of the common challenges in generating optical tweezers is the spherical aberration generated by the glass thickness of the vacuum cell (Fig.~\ref{Fig1}a). We demonstrate that when we run our automation algorithm using a deformable mirror, it can correct this aberration upto a glass thickness of $1\mm$. We begin with a description of the experiment.\\

\begin{figure*}
    \centering
    \includegraphics[scale=1]{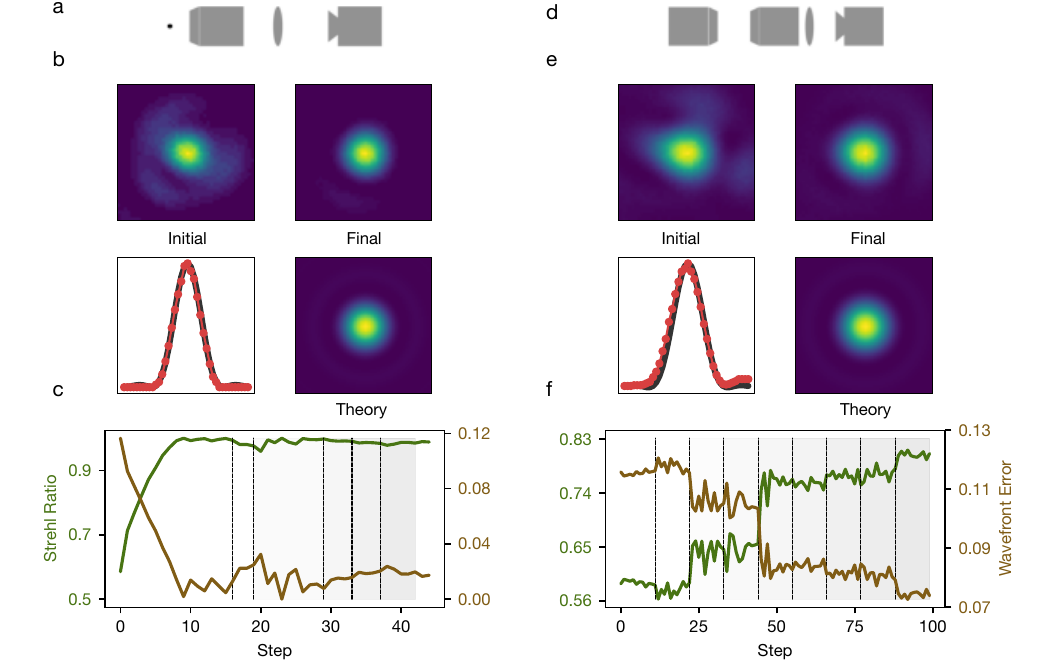}
    \caption{\textbf{Automated mechanical alignment:}\textbf{a} Schematic of the set up used to image a PSF, corresponding to data presented in \textbf{b} and \textbf{c}. \textbf{b} Image of the PSF before and after optimization. The bottom left panel shows a cross section of the PSF comparing theory (black) and experiment (red). \textbf{c} Shrehl ratio and wavefront error, during the running of the modified AM algorithm. The dashed lines separate the optimizations corresponding to different $\vec{v}_i$. We use $\ell = 9$ vectors. The maximum Strehl ratio reached is close to $1$. \textbf{d} Schematic of the set up used to produce and image a tweezer, corresponding to data presented in \textbf{e} and \textbf{f}. \textbf{e} Image of a single tweezer before and aftyer alignment. \textbf{d} Shrehl ratio and wavefront error, during the running of the modified AM algorithm while optimizing a tweezer. The maximum Strehl ratio reached is above $0.8$. The step-like features in the Strehl ratio and the wavefront error correspond to those $\vec{v}_i$'s that are observed to be effective in the optimization procedure. We use $\ell=22$ for this case (see text). }
    \label{tweezer}
\end{figure*}

\tocless\section{Experimental setup }
\begin{figure*}[ht]
    \includegraphics[scale=1]{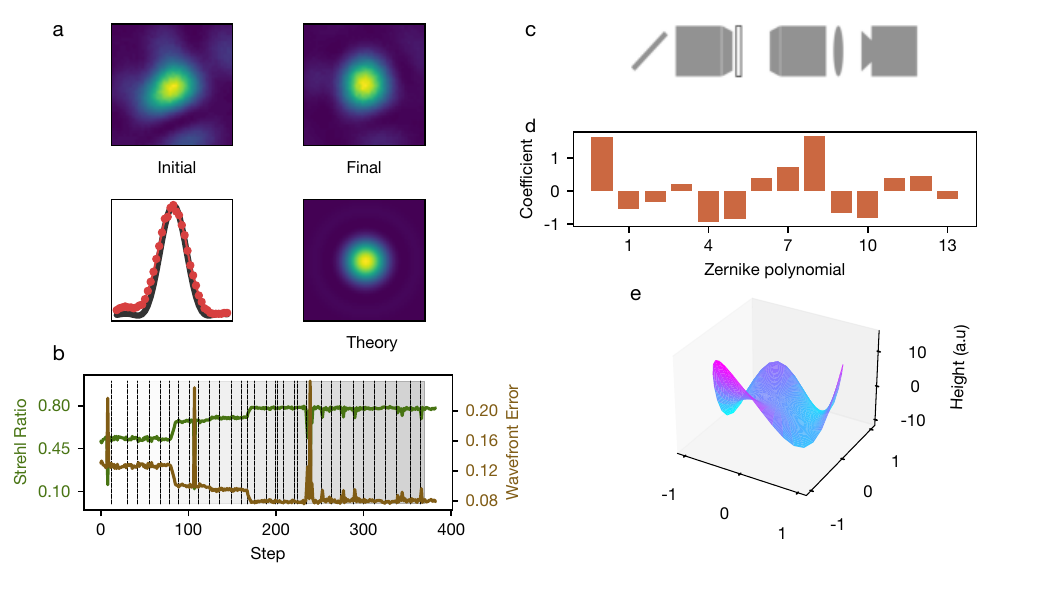}
    \caption{ \textbf{Tweezer optimization with a deformable mirror:}  \textbf{a} Image of the tweezer before and after optimization and comparison with theory. \textbf{b} The Strehl ratio and the wavefront error through the optimization. As before, the dashed lines separate different $\vec{v}_i$'s. We use $\ell= 38$ vectors. Note the multiple step-like features. They correspond to the vectors $\vec{v}_i$ that are observed to be effective at optimizing the Strehl ratio. In particular, the feature seen at $\sim 80$ steps in corresponds to optimization of one of the parameters of the deformable mirror which corrects for the aberration caused by the glass thickness. \textbf{c} schematic of the setup used. The mirror at $45^{\circ}$ is the deformable mirror. A glass window of thickness $1\mm$ is placed after the first objective. \textbf{d} Zernike coefficients of the final shape of the mirror. \textbf{e} final shape of the mirror. The unit on the $z-$axis is Volts, where each $V$ corresponds to a physical displacement of $\sim 30 \nm$. This data was taken with $520 \nm$ laser. }
    \label{tweezer_dm}
\end{figure*}

We utilize two experimental configurations: one for imaging a point source and the other for generating a tweezer. The former involves a single objective focused on a point source (see Fig.\ref{tweezer}a), while the latter employs two objectives facing each other (see Fig.\ref{tweezer}d). To investigate the optimization of the deformable mirror alongside the orientation of the objectives, we insert a glass window with a thickness of $1\mm$ after one of the objectives and place a deformable mirror in the beam path (refer to Fig.~\ref{Fig1}a, Fig.\ref{tweezer_dm}c and Fig.\ref{PSF_dm}c for schematics).

In the first configuration, we use a pinhole with a diameter of $250 \nm$, illuminated by a laser with a wavelength of $635 \nm$, as the point source. The pinhole is positioned near the focus of an objective with NA$=0.67$ and EFL$=4 \mm$, imaging onto a camera with a tube lens of focal length $200 \mm$. This setup results in the magnified ($50$X) image of a point source. The objective is mounted on a translation stage controlled by a piezoelectric inertia actuator, allowing us to adjust the distance $z$ between the pinhole and the objective. Furthermore, the tip ($\theta$) and tilt ($\phi$) of the objective can be tuned using a piezoelectric inertia actuator-controlled mirror mount.

In the second variant of this setup, a deformable mirror is introduced before imaging. We use a MEMS deformable mirror, allowing control over $\theta$, $\phi$, $z$, and the mirror's shape. The latter is characterized using the first $14$ non-trivial Zernike coefficients, corresponding to Zernike polynomials $Z_{nm}$ with $n=1, 2, 3, 4$. For technical details of the experimental components, refer to ref.~\cite{supplements}.

The second configuration consists of two identical objectives, both with with NA$=0.67$, EFL=$4 \mm$ and a back pupil of $5.4 \mm$ (Fig.~\ref{tweezer}d). The objectives are placed facing each other, and a collimated laser beam in incident on the back pupil of one of them, producing a tweezer at the focus. We use two wavelengths, $520 \nm$ and $405 \nm$. We use a tube lens after the second one to image the tweezer on a camera, with a magnification of $50$X. In this configuration, both the objectives have three controllable parameters each, $\theta_1, \phi_1, z_1, \theta_2, \phi_2$ and $z_2$. When using the deformable mirror in this configuration, we have additional  $14$ variables. 

Both the point source and the tweezer are imaged on a camera. We use a camera with a pixel size of $4.5\um $ when imaging the PSF and a different camera with a pixel size of $1.85 \um$ when imaging a tweezer. We use the image produced on the camera to estimate the strehl ratio and the wavefront error. We follow the procedure described in ref.~\cite{David_thesis} to estimate the Strehl ratio. The details of estimating the wavefron error are presented in ref.~\cite{supplements}. The Strehl ratio is a function of all of the controllable parameters. The problem in both cases is to optimize the controllable parameters in order to maximize the Strehl ratio or minimize the wavefront error. Below, we describe the algorithm we develop for the optimization.  \\

\tocless\section{The modified AM algorithm}
This algorithm is inspired by the standard ``cross-walking" used to couple lasers into fibers and mathematically, it is a generalization of the well known AM algorithm in convex optimization~\cite{GRIPPO2000127}. 

Cross walking is a standard procedure in optical alignment.  For instance, when one uses two mirrors with tips $\theta_1, \theta_2$ and tilts $\phi_1, \phi_2$ to optimize the coupling of laser into a fiber tip, optimizing the four variables independently is almost always insufficient. We follow it up by ``walking" the pairs $\theta_1, \theta_2$ and $\phi_1, \phi_2$ and alternating between them. Formally, the walking can be described as optimizing the output power $P(\theta_1, \theta_2, \phi_1, \phi_2)$ over $a\theta_1+b\theta_2$ and over $c\phi_1 + d\phi_2$ for appropriately chosen $a, b, c, d$. Geometrically, one can consider this as optimizing along the linear axes in the $4-$dimensional space spanned by $\theta_1, \theta_2, \phi_1, \phi_2$, defined by the direction $(a, b, 0, 0)$ and $(0,0, c, d)$. The independent optimization over the four variables can be considered as optimizing along the linear axes $(1,0,0,0), (0,1,0,0), (0,0,1,0)$ and $(0,0,0,1)$. 

The independent optimization over the four variables can be connected to the standard alternating minimization (AM) algorithm~\cite{author1984information}. If $f(\vec{x})$ , with $\vec{x}=(x, y)$ is a $2-$variable function which we intend to optimize, under the AM algorithm, we optimize over each variable $x, y$. Formally, it can be written as  
\begin{equation}
    \begin{split}
        (x_{2k+1}, y_{2k+1})&= \text{argmin}_{x} f(x, y_{2k})\\
        (x_{2k+2}, y_{2k+2})&= \text{argmin}_{y} f(x_{2k+1}, y)\\
    \end{split}
\end{equation}
Here, $k$ is the iterator. If the function $f$ satisfies a set of conditions known as the \textit{five-point property,} this algorithm converges to the global minima. See ref.~\cite{charles} for a review and ref.~\cite{andresen2016convergence} for details on the convergence. The AM algorithm belongs to a larger class of sequential algorithms known as sequential unconstrained minimization algorithms (SUMMA)~\cite{byrne2008sequential}. It can be connected to the expectation maximization algorithm used in machine learning~\cite{143375, Haas2002TheEA}.

It is the experience of every experimentalist that one can't couple a laser beam into a fiber only using the alternating algorithm. One has to cross walk the beams. This is perhaps because the function being optimized doesn't always satisfy the five-point property (see Fig.~\ref{Fig1}c for a reasoning).  The cross-walking can be considered as a modification of the above algorithm with three steps in each iteration, where we also optimize over a third axis, $ax+by$ besides $x$ and $y$:
\begin{equation}
    \begin{split}
        (x_{3k+1}, y_{3k+1})&= \text{argmin}_{x} f(x, y_{3k})\\
        (x_{3k+2}, y_{3k+2})&= \text{argmin}_{y} f(x_{3k+1}, y)\\
        (x_{3k+3}, y_{3k+3})&= \text{argmin}_{\lambda} f(x_{3k+2}+a\lambda, y_{3k+2}+b\lambda)\\
    \end{split}
\end{equation}
There is no reason to stop at one new direction $ax+by$; we can add more directions within the iteration. Figs.~\ref{Fig1}b,c show a numerical example of the modified AM applied to a two variable function, with four steps in each iteration, optimization over $x, y, x+y$ and $x-y$.

We now formalize the definition of modified AM algorithm for a function $f(\vec{x})$ of $\vec{x}=(x_1, \cdots, x_m) \in \mathbbm{R}^m$ variables. We pick fixed vectors $\vec{v}_1, \cdots, \vec{v}_{\ell} \in \mathbbm{R}^m$. The $k$-th iteration consists of $\ell$ steps:
\begin{equation}\label{AM}
    \begin{split}
        \vec{x}_{\ell k+1}&= \text{argmin}_{\lambda} f(\vec{x}_{\ell k}+\lambda\vec{v}_1)\\
        \vec{x}_{\ell k+2}&= \text{argmin}_{\lambda} f(\vec{x}_{\ell k+1}+\lambda\vec{v}_2)\\
         & \vdots\\
        \vec{x}_{\ell k+n}&= \text{argmin}_{\lambda} f(\vec{x}_{\ell k+\ell-1}+\lambda\vec{v}_{\ell})\\
    \end{split}
\end{equation}
It is intuitive that $\vec{v}_1, \cdots, \vec{v}_{\ell}$ should span all of $\mathbbm{R}^m$. Moreover, it helps if they \textit{overspan} the space. Indeed, one can include all the basis elements $(1, 0, 0, \cdots, 0), \cdots, (0, \cdots, 0, 1)$ in this set. The effectiveness of cross-walking indicates that having more vectors, in other directions helps. The choice of the set should depend in general on our understanding of the correlations that maybe present between the variables. \\

\tocless\subsection{Optimization algorithm} 
We now describe the algorithm used to optimize the function within each step. This is a variant of gradient descent. The function $f(\vec{x}_{{\ell}k+i}+\lambda\vec{v}_{i+1})$ is the Strehl ratio measured experimentally by averaging over $30$ images taken by the camera (see ref.~\cite{supplements} for details). And the variables $\vec{x}$ include the orientation of the objectives and the shape of the deformable mirror.  In order to optimize this over $\lambda$, we begin with an initial value $\lambda_1$ and we use the following equation update after each measurement of the average Sterhl ratio:
\begin{equation}\label{update}
    \lambda_{j+1} = \lambda_j + \gamma \Delta f_j
\end{equation}
Here, $\Delta f_j = (f(\vec{x}_{nk+i}+\lambda_{j}\vec{v}_{i+1}) - f(\vec{x}_{nk+i}+\lambda_{j-1}\vec{v}_{i+1}))$ is the difference between the measured Strehl ratios before and after the $j-$th step. We use $\lambda_0=0$. $\gamma$ is a feedback parameter. This iteration is continued till $j=N$ or till $\Delta f_j < \epsilon_{thresh}$ (a threshold value), whichever comes first. The threshold is set to be equal to the standard deviation of the mean of the Strehl ratio, over the $30$ images taken. See ref.~\cite{supplements} for more details on how $\lambda_1, \gamma, \epsilon_{thresh}$ and $N$ are chosen, and how $\gamma$ is updated incase of overcorrection. In fact, one could divide the $\Delta f_j$ in Eq.~\ref{update} by $\lambda_j -\lambda_{j-1}$ to reduce the algorithm to the standard gradient descent.  However,  $\lambda_j -\lambda_{j-1}$ may have a large error. For instance this happens when $\lambda_j$ is set by open loop piezoelectric inertia actuators. This results in a failure to converge, if we use the standard gradient descent algorithm. Besides the control parameters, there are also errors in $f$, i.e., the measured Strehl ratio. The experimental data shows that our algorithm is robust to both of these error. See ref.~\cite{supplements} for an analysis of these errors. We now present the experimental results.\\

\begin{figure}[ht]
    \centering
    \includegraphics[scale=1]{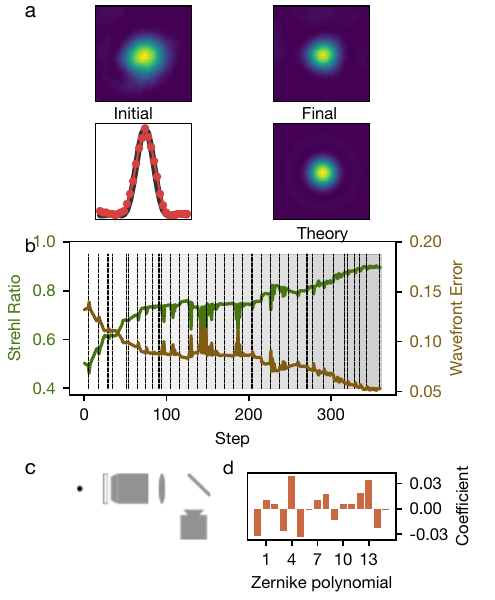}
    \caption{ \textbf{PSF Optimization with a deformable mirror:} \textbf{a} Image of the PSF before and after optimization while imaging a point source, with a $1\mm$ glass window between the source and the objective. The top left panel shows the initial PSF after coarse alignment. Note the enlarged diameter of the PSF, caused by the spherical aberrations due to the glass window. The top right panel shows the PSF after running the algorithm. The deformable mirror fixes the spherical aberration. \textbf{b} Shrehl ratio and wavefront error, during the optimization. Once again, the dashed lines separate the optimizations corresponding to different $\vec{v}_i$. We use $\ell=25$. The mechanical alignments correspond to steps $1-105$ on the x-axis. Note the plateau reached by $100$ steps --- this corresponds to the best possible alignment without correcting for the spherical aberrations using the deformable mirror. The increase at $200$ corresponds to the correction of spherical aberrations. \textbf{c} schematic of the setup used. The mirror at $45^{\circ}$ is the deformable mirror. \textbf{d} Zernike coefficients of the shape of the mirror post optimization. This data was taken with a $635 \nm$ laser.  }
    \label{PSF_dm}
\end{figure}

\tocless\section{Alignment without the deformable mirror}
We first present the results for the mechanical alignment of the objectives without the deformable mirror. While imaging a point source, we have three variables $\theta, \phi, z$. We use $\ell =9$ vectors 
\begin{equation}\label{one_objective}
    \begin{split}
        \vec{v}_1 &= (1,0,0), \ \vec{v}_2 = (0,1,0), \ \vec{v}_3 = (0,0,1)\\
        \vec{v}_4 &= (0,1,1), \ \vec{v}_5 = (0,1,-1), \ \vec{v}_6 = (1,0,1)\\
        \vec{v}_7 &= (-1,0,1), \ \vec{v}_8 = (1,1,0), \ \vec{v}_9 = (1,-1,0)\\
    \end{split}
\end{equation}
The fist three vectors correspond to independent optimization of the three variables. The rest of them correspond to cross walking pairs of variables. While imaging a tweezer, we have six variables $\theta_1, \phi_1, z_1, \theta_2, \phi_2, z_2$ three each for the two objectives. We use $\ell =22$. The first $9$ are the same as before:
\begin{equation}\label{two_objectives1}
    \begin{split}
        \vec{v}_1 &= (1,0,0, 0, 0, 0), \ \vec{v}_2 = (0,1,0, 0, 0, 0)\\
        \vec{v}_3 &= (0,0,1, 0, 0, 0), \ \vec{v}_4 = (0,1,1, 0, 0, 0)\\
        \vec{v}_5 &= (0,1,-1, 0, 0, 0), \ \vec{v}_6 = (1,0,1, 0, 0, 0)\\
        \vec{v}_7 &= (-1,0,1, 0, 0, 0), \ \vec{v}_8 = (1,1,0, 0, 0, 0)\\
        \vec{v}_9 &= (1,-1,0, 0, 0, 0)\\
    \end{split}
\end{equation}
The next $9$ vectors are the analogous ones for the second objective
\begin{equation}\label{two_objectives2}
    \begin{split}
        \vec{v}_{10} &= (0,0,0,1,0,0), \vec{v}_{11} = (0,0,0,0,1,0)\\
        \vec{v}_{12} &= (0,0,0,0,0,1), \vec{v}_{13} = (0,0,0,0,1,1)\\
        \vec{v}_{14} &= (0,0,0,0,1,-1), \vec{v}_{15} = (0,0,0,1,0,1)\\
        \vec{v}_{16} &= (0,0,0,-1,0,1), \vec{v}_{17} = (0,0,0,1,1,0)\\
        \vec{v}_{18} &= (0,0,0,1,-1,0)\\
    \end{split}
\end{equation}
The last $4$ variable correspond to cross walking between the tip and tilt of the two objectives.
\begin{equation}\label{two_objectives3}
    \begin{split}
        \vec{v}_{19} &= (0,1,0, 0, 1, 0), \ \vec{v}_{20} = (0,1,0, 0, -1, 0)\\
        \vec{v}_{21} &= (0,0,1, 0, 0, 1), \ \vec{v}_{22} = (0,0,1, 0, 0, -1)\\
    \end{split}
\end{equation}

We begin the experiment with a coarse alignment. We align the laser beam to the center of the camera sensor and coarse align the objective(s) so as to maintain the position of the imaged laser spot on the camera sensor. Fig.~\ref{tweezer}b, top-left panel shows the initial image of the PSF after the coarse alignment. We then run the AM algorithm, following Eq.~\ref{AM}. Fig.~\ref{tweezer}c shows the Strehl ratio and the wavefront error through the runtime of the algorithm. The shades of grey separated by the dashed lines represent different steps within the iteration. The Strehl ratio reaches a maximum value close to $1$, after $5-10$ minutes of optimization. We use one iteration, i.e., $k$ in Eq.~\ref{AM} takes just one value.  Fig.~\ref{tweezer}b top-right panel shows the image on the camera after alignment, which maybe compared with the expected PSF, shown in Fig.~\ref{tweezer}b bottom-right. The bottom-left shows this comparison for a cross section. 

Fig.~\ref{tweezer}d-f shows similar results for a single tweezer. We use $\ell=22$ for this case (see text). The step-like features observed correspond to the various $\vec{v}_i$s that had a significant impact in the optimization. We reach a Strehl ratio of $\approx 0.82$ within $15$ minutes of optimization. We attribute the residual error to the standard, non-low wavefront error mirrors used in the experiment. \\

`

\tocless\section{Alignment with the deformable mirror}
We model the shape $S$ of a deformable mirror using the first $14$, non trivial Zernike polynomials corresponding to $n=1, 2, 3$ and $4$:
\begin{equation}
    S(r, \phi) =\sum_{n=1}^4\sum_{m=-n}^n s_{nm}Z_{nm}(r, \phi)
\end{equation}
Here, $r, \phi$ are the position variables on the mirror and $S(r, \phi)$ is it's height deformation at the point $(r, \phi)$. The parameters $s_{nm}$ can be controlled on the mirror and are optimized along with the mechanical variables of the objective(s). The standard control provided on a MEMS mirror and a piezo deformable mirror (DMP40) are not the Zernike coefficients. See ref.~\cite{supplements} for a calibration of both the mirrors and details of the mapping between the control parameters and the Zernike coefficients.  

When optimizing the imaging of a point source, we use $\ell = 25$. This includes the $9$ vectors corresponding to the position variables of the objective (Eq.~(\ref{one_objective})), the $14$ Zernike coefficients and finally the two cross walking vectors between the defocus, $s_{20}$ and the focus $z$ of the objective. Fig.~\ref{PSF_dm} shows the results of this optimization, with a $1 \mm$ glass window. We can clearly see the effect of the spherical aberrations in the PSF before optimization. In Fig.~\ref{PSF_dm}b, we can see the optimization of the Zernike coefficients correcting this aberrations, around step $200$. We reach a Strehl ratio of $\approx 0.9$ in about one hour. 

When optimizing tweezer generation, we use $\ell = 38$. This includes the $22$ vectors corresponding to the two objectives, discussed in Eqs.~\ref{two_objectives1}, ~\ref{two_objectives2}, ~\ref{two_objectives3}, the $14$ Zernike coefficients and two cross walkings between $s_{20}$ and $z_1$. The results are shown in Fig.~\ref{tweezer_dm}. Once again, we use a $1\mm$ window between the objectives (Fig.~\ref{tweezer_dm}c). We reach a Strehl ratio of $\approx 0.8$ in about $90$ minutes.  \\

\tocless\section{Conclusions}
We have developed and tested a new algorithm to automate fine-tuned optical alignment and optimize parameters of a deformable mirror. We applied this algorithm to generate a high quality tweezer and demonstrate diffraction limited imaging. Our results greatly simplify the buildup of high quality tweezers and diffraction limited imaging. This build up usually involves specialized low-wavefront error optics (e.g mirrors), low stress optical mounts and a  delicate alignment procedure~\cite{David_thesis}. We have shown that using our algorithm, diffraction limited imaging and tweezers can be generated with non-specialized optics and mounts and starting from a rudimentary coarse alignment done manually. The experiments were performed on a non-floating table with no vibration isolation, demonstrating the robustness of our algorithm. Moreover, this automated alignment technique can be used to modularize the experimental system into replaceable and interchangeable parts. For instance, the glass cell can be a replaceable module, where one uses the automated alignment to realign the objectives after replacing the cell. 

Our automation algorithm gives a standardized alignment procedure and therefore can be used to compare the quality  of the tweezers and imaging when different optical elements such a dichroic mirrors, SLMs, AODs are inserted into the beam path. A possible future direction is to use the algorithm to optimize other objective functions, which maybe of interest when one generated several tweezer beams. One can also explore the performance of the algorithm to optimize the parameters of other optical devices such as SLMs, DMDs. \\

\section*{Acknowledgements}
We thank Michael J. Martin for useful discussions. Research presented in this article was supported by the Laboratory Directed Research and Development program of Los Alamos National Laboratory under project number(s) 20230779PRD1, 20210301ER and 20230188DR. 

\paragraph*{\textbf{Code Availability}} The code developed here will be shared upon a reasonable request. 

\paragraph*{\textbf{Competing interests}} The authors declare no competing interests.

\bibliography{References}
\appendix

\cleardoublepage

\setcounter{figure}{0}
\setcounter{page}{1}
\setcounter{equation}{0}
\setcounter{section}{0}

\renewcommand{\thepage}{S\arabic{page}}
\renewcommand{\thesection}{S\arabic{section}}
\renewcommand{\theequation}{S\arabic{equation}}
\renewcommand{\thefigure}{S\arabic{figure}}
\onecolumngrid
\begin{center}
\huge{Supplementary Information}
\vspace{5mm}
\end{center}
\twocolumngrid
\normalsize
\tableofcontents

\section{Devices and calibrations}
\begin{figure*}
    \centering
    \includegraphics[scale=0.4]{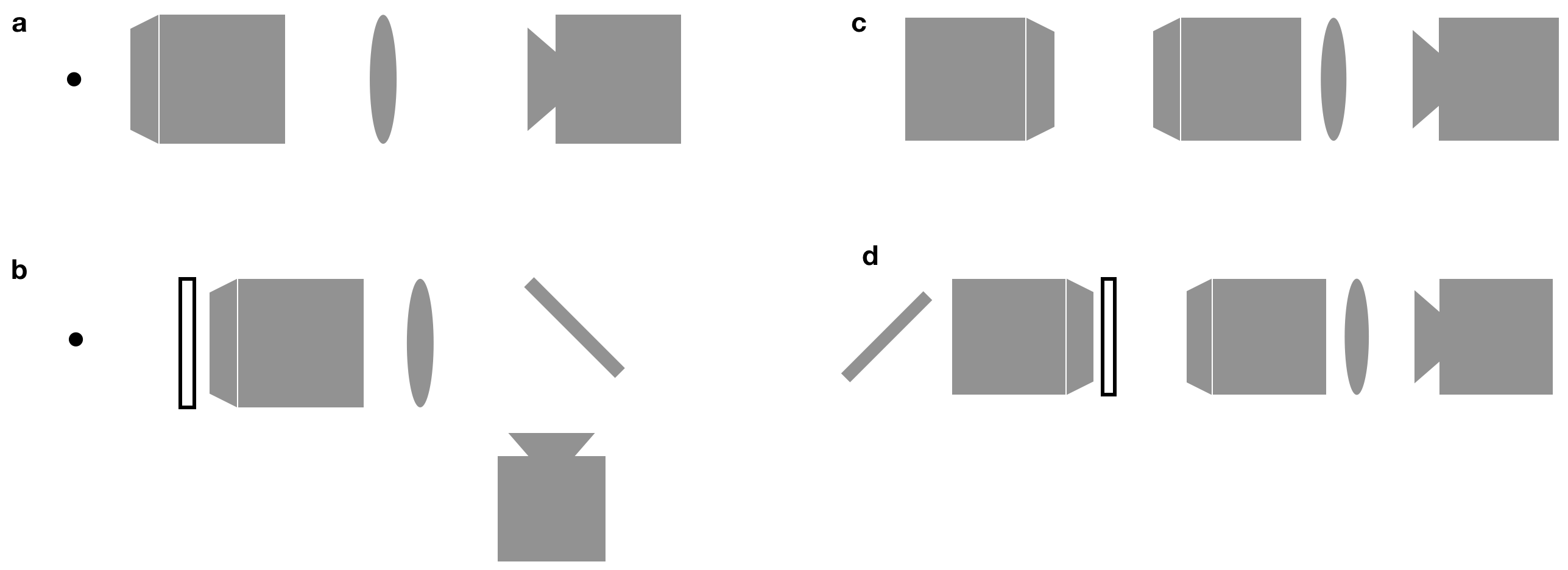}
    \caption{\textbf{Experimental setup:} The four settings, \textbf{a}, \textbf{b}  to observe the point spread function and \textbf{c}, \textbf{d} to produce a tweezer. \textbf{a} and \textbf{c} don't use a deformable mirror while \textbf{b} and \textbf{d} do. }
    \label{FigS1}
\end{figure*}

We use four settings, shown in Fig.~\ref{FigS1}. The devices used in the setup are listed in table~\ref{Table}. For setting Fig.~\ref{FigS1}b, we use the MEMS deformable mirror. This is a membrane based array of micromirrors, with a pitch of $3\um$ . It has $140$ pixels, arranged in a $12\times 12$ array with the corners removed. For setting Fig.~\ref{FigS1}d, we use a piezo based DMP40 mirror. This mirror is bigger, with a $10 \mm$ active area but has $40$ deformable units. See sec.~\ref{DM} for details on calibration of these two mirrors. In the following subsection, we outline our calibration procedure for the piezoelectric inertia actuators.

\begin{table*}[]
    \centering
    \begin{tabular}{|c|c|c|}
    \hline
    \textbf{Component} & \textbf{Part no.} & \textbf{Specs} \\
    \hline
    Pinhole & Technologie Manufaktur, TC-RT01 & Smallest hole: $250$\nm \\
    \hline
        Objective & OptoSigma, PAL-50-NIR-HR-LC07 & NA=0.67, EFL=4 \mm, Pupil=5.4 \mm\\
        \hline
        Piezoelectric inertia actuator & Newport 8821, 8302 & openloop \\
        \hline
        Camera 1& Allied vision Alvium & $1.85 \um$  pixels \\
        \hline
        Camera 2& Teledyne FLIR BFS-U3-88S6M-C & $4.5\um $pixels. \\
        \hline
        Deformable mirror 1& Boston micromachnines, MEMS& $140$ micromirrors \\
        \hline
        Deformable mirror 2&  Thorlabs, DMP40& $40$ piezos  \\
        \hline
    \end{tabular}
    \caption{Devices and their specifications}
    \label{Table}
\end{table*}

\begin{figure}[ht!]
    \includegraphics{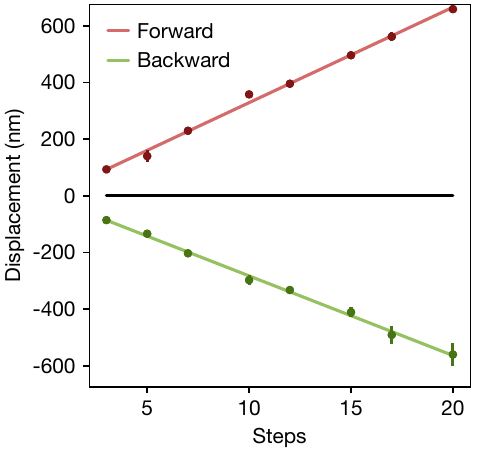}
    \caption{\textbf{Piezoelectric inertia actuator calibration:} The transverse displacement of the PSF, measured using the imaging objective, due to a transverse piezoelectric inertia actuator. This is used to calibrate the piezoelectric inertia actuator. The circular markers are the data points and the straight line is the linear fit. The error bars are statistical. There is a systematic difference between the forward and the backward motion of the piezoelectric inertia actuator. The fit parameters are $33$\nm\  and $28$ \nm\  per step respectively.   }
    \label{FigS2}
\end{figure}

\subsection{Calibrating the piezoelectric inertia actuators}
We calibrate the piezoelectric inertia actuators using the setting Fig.~\ref{FigS1}a. We use an additional Newport $8302$ on the pinhole mount within the focal plane. We then move the pinhole by a fixed number of steps $s$ and use the imaged PSF to measure the resulting displacement in the foal plane. We take $30$ data points for each value of $s$. Fig.~\ref{FigS2} shows the measured average displacement with the standard deviation (i.e., the errorbars) for various values of $s$. Positive (i.e., forward) and negative (i.e., backward) values of $s$ are shown and analyzed separately, since we expect a systematic difference between them. Note that although the device is open loop, the displacement is relatively consistent. The fit values are $33 \nm$ per step for forward movement and $28\nm$ per step for backward movement.

\subsection{Measuring the Strehl ratio}
\begin{figure}
    \centering
    \includegraphics{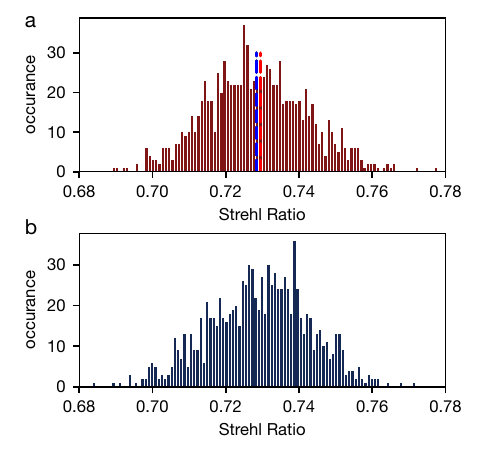}
    \caption{\textbf{Fluctuations in the Strehl ratio:} \textbf{a} Experimental data of $1000$ samples of the Strehl ratio. The dashed lines represent the mean value (black) and the $p-$mean for $p=2, 10$ (blue, red). The standard deviation is about $0.013$. \textbf{b} Simulation of the fluctuation in the Strehl ratio, where the distribution of the wavefront error is modeled by a Gaussian.  }
    \label{FigS6}
\end{figure}

We estimate the Strehl ratio from the PSF using the formula:
\begin{equation}
    SR= \frac{I_{expt}(\textbf{0}) \int I_{th}(\textbf{r}) d^2\textbf{r}}{I_{th}(\textbf{0}) \int I_{expt}(\textbf{r}) d^2\textbf{r}}
\end{equation}
Here, $I_{th}(\textbf{r})$ ($I_{expt}(\textbf{r})$) is the theoretical (experimental) intensity pattern. The theoretical pattern is given by 
\begin{equation}
    I_{th}(\textbf{r}) = \frac{4 J_1^2(2\pi r NA \times M/\lambda)}{r^2}
\end{equation}
Here, $J_1$ is the Bessel function with $n=1$,  $M$ is the magnification ($M=50$ in most of our experiments) and $NA$ is the numerical aperture. See ref.~\cite{David_thesis} for more details. The experiment is setup on a non-floating, regular office table,which is not vibration isolated. Moreover, the setp is in a noisy enviroment (i.e., not a well isolated laboratory). Consequently, the measured Strehl ratio fluctuates. In Fig.~\ref{FigS6}, we show an experimental dataset of the statistics of the Strehl ratio computed using $1000$ samples of the PSF, taken with an exposure of $23 \ms$ each, with a gap of $100 \ms$ after each sample. The Distribution can be modeled by a normally distributed wavefront error, as shown in Fig.~\ref{FigS6}b. Therefore, in order to make the algorithm robust to this fluctuation, we take $30$ images and compute the average Strehl ratio. The threshold value $\epsilon_{thresh}$ (see main text) is given by the standard deviation of the mean of the Strehl ratio over these $30$ measurements. Typically, this is $\epsilon_{thresh} \approx 0.002$. One can take more than $30$ images, and reduce the threshold further. However, the algorithm will take longer to run in this case and will pick up additional errors and drifts and it maybe counter productive.   

\subsection{Calibrating the deformable mirrors}\label{DM}
\begin{figure*}
    \centering
    \includegraphics{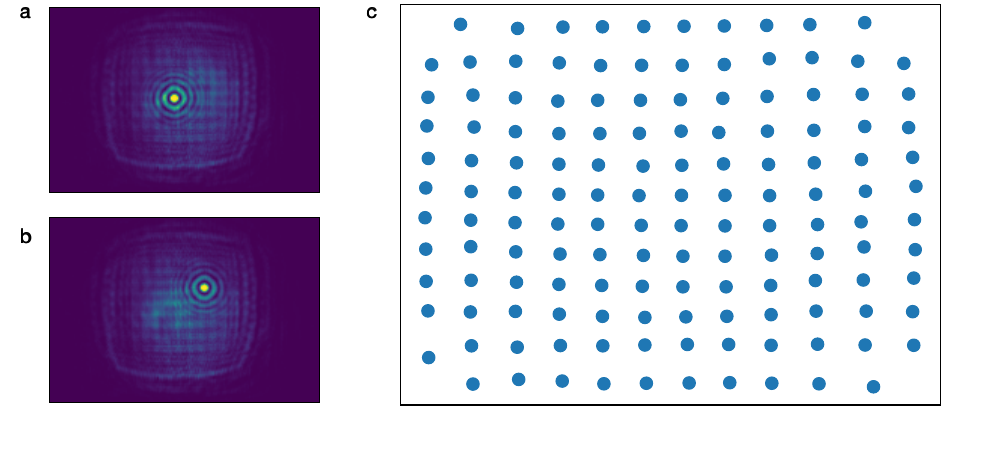}
    \caption{\textbf{MEMS Calibration:} \textbf{a, b} Images of a collimated beam with a particular pixel in the MEMS device is poked. The Location of the poked pixel can be calibrated by identifying the peak of the intensity. \textbf{c} The calibrated locations of all the $140$ pixels using this technique. }
    \label{FigS3}
\end{figure*}

\textbf{MEMS mirror:} In the MEMS mirror, the $140$ pixels are indexed from $0$ to $139$ in some order.  We calibrate the physical positions corresponding to the indices using a feature called ``poking", available in the device. This action moves a pixel with a chosen index, keeping the rest of them in the zero position. If the mirror surface is illuminated by a collimated beam and imaged on the camera, the poking results in an Airy-like pattern at the physical location of the pixel (see Fig.~\ref{FigS3} a, b). We use this to calibrate the physical positions of the pixels (Fig.~\ref{FigS3}c).

\begin{figure}
    \centering
    \includegraphics{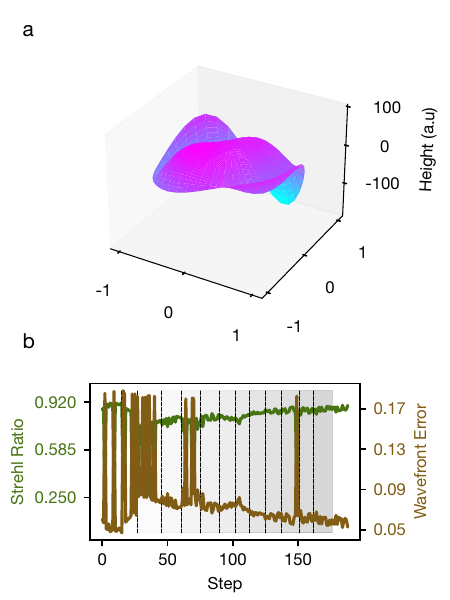}
    \caption{\textbf{Calibrating DMP40:} \textbf{a} The optimal input shape after \textit{flattening} the mirror (see text). \textbf{b} The Strehl-Ratio and wave front error during the flattening of the mirror.  }
    \label{FigS5}
\end{figure}

\textbf{DMP40: }This mirror is controlled by $40$ voltages, ranging from $0$ to $200 V$, applied to the piezos. The mirror needs to be \textit{flattened} before use, i.e., we need to find the optimal volatages at which the mirror is flat. Setting all the voltages to $100 V$ should, theoretically, flatten the mirror. However, we found some residual errors, which we calibrated by optimizing the focus of a collimated laser beam, reflected off the mirror and  focused on the camera, without the objectives (Fig.~\ref{FigS5}). 

\begin{figure}[h]
    \centering
    \includegraphics{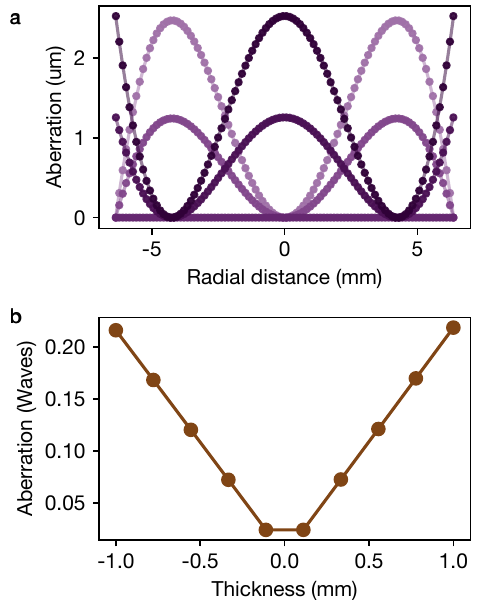}
    \caption{\textbf{Numerical calculation of the aberrations:} \textbf{a} Numerical computation of the aberration due to a glass thickness. The five curves correspond to thickness $-2, -1, 0, 1$ and $2$\mm. The darkest curve corresponds to $2\mm$ and the lightest corresponds to $-2\mm$. The circular markers represent numerical computations and the solid curve corresponds to the best-fit, using Zernike polynomials upto $n=4$. \textbf{b} The magnitude of aberration for various thicknesses, for $\lambda =520 \nm$.  }
    \label{FigS4}
\end{figure}

\section{Spherical Aberration due to glass thickness:} We show a numerical simulation of the expected aberration of the wavefront due to a glass thickness, assuming ideal properties of the objective in Fig.~\ref{FigS4}. Note that the pitch is about $2.5 \um $, for a thickness of $2$\mm, which is less than the pitch of the two deformable mirrors. Therefore, we should be able to correct these aberrations for thicknesses up to $2\mm$, in theory.

\section{Main text figures}
\subsection{Main text figure 2}
In Fig.~$2$a-c, we use red laser ($635$ \nm) to illuminate a $250$\nm pinhole and imagie it using the objective. This experimental setup does not involve a deformable mirror and therefore, the modified AM algorithm  is used to optimize the three variables corresponding to the focus, tip and tilt of the objective. In Fig.~$2$d-f, we use a green laser ($520$ \nm) to produce a single tweezer. 
\subsection{Main text figure 3}
The data in this figure was taken with the green laser ($520$ \nm), and the DMP40 deformable mirror. The mirror has a diameter of $10$\mm, while the back pupil of the objective is $5.4$\mm. We therefore used a $2X$ telescope between the mirror and the objective. 
\subsection{Main text figure 4}
This data was taken with the red laser ($635$ \nm) and the MEMS mirror. The latter has a physical dimension of $5 $\mm $\times $ $5$\mm and therefore, the effective aperture of the objective is reduced by $10\%$. The NA is therefore reduced to $0.55$. The data is evaluated with this NA.

\section{Code description}
All the devices, including the camera, piezoelectric inertia actuators and the deformable mirrors are controlled via API on python. We found that some of the APIs are incompatible with python $3.8$. We use python $3.10$. Central to the code is the python object \textit{State}, with represents the state of the setup within the optimization procedure. It includes all the relevant attributes such as the current PSF, the current Strehl ratio, the current shape of the deformable mirror, the current vector $\vec{v}$ and the current value of the feedback parameter $\gamma$. During every ``update", a new tweezer or PSF image is taken and the piezoelectric inertia actuator and the deformable mirror's states are changed according top Eq.~$4$ of the maintext. 

\begin{figure}
    \centering
    \includegraphics{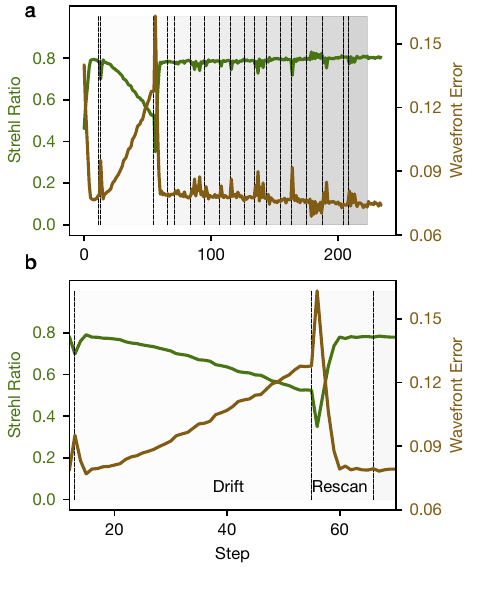}
    \caption{\textbf{Drift correction:} \textbf{a} Strehl ratio and wave-front error during the optimization. After two steps in the optimization (i.e., two $\vec{v}$'s), we can see a steady decrease in the Strehl ratio, caused by a focal drift. The code detects such a drift, and responds by a rescan of the focus before continuing. \textbf{b} zoomed in version of the plot. }
    \label{FigS7}
\end{figure}

The algorithm is often disturbed by various experimental effects such as drifts, and errors of external origin.  We list a few protective scripts in the code to counter these effects:

\begin{itemize}
    \item \textbf{Protection against over-correction:} The value of $\gamma$ is chosen arbitrarily at the beginning. If it is too large, it leads to over correction and consequently, the Strehl ration oscillates rather than converging to the optimum. The code detects if the Strehl ratio is oscillating and it reduces $\gamma$ by a factor of $2$.
    \item \textbf{Protection against drifts:} After optimizing for each vector $\vec{v}$, the code checks if the Strehl ratio post optimization is significantly less than before the optimization. If so, the code interrupts the process and aligns the focus, assuming that the focus has drifted. We have observed that almost all cases of drifts are caused by focal shifts. See Fig.~\ref{FigS7} for an example.
    \item \textbf{Protection against externalities:} Sometimes, the Strehl ratio could show a large dip, due to an external reason, unrelated to the variables being optimized within the algorithm (e.g. door slam). This will lead to the code deciding that the Strehl ratio is strongly sensitive to the variable being optimized, but the optima is in the opposite direction. As a result, it will make a big shift in the variable in the opposite direction. To avoid this, in cases where the sign of $\lambda$ switches, we bound the change by the magnitude of the previous change. That is, 
    \begin{equation}
        \lambda_{j+1}-\lambda_j  =- \max\{|\gamma \Delta f_j|,|\lambda_{j}-\lambda_{j-1}| \}
    \end{equation}
    whenever $\Delta f_j <0$ (see Eq.~$4$ of main text).
\end{itemize}

\end{document}